\documentclass[twocolumn,pra,showpacs,superscriptaddress]{revtex4}

\usepackage{graphicx}
\usepackage{dcolumn}
\usepackage{bm}

\begin{document}

\title{Macroscopic Klein Tunneling in spin-orbit coupled Bose-Einstein Condensates}
\author{Dan-Wei Zhang}
\affiliation{Laboratory of Quantum Information Technology and
SPTE, South China Normal University, Guangzhou, China}
\affiliation{Department of Physics and Center of Theoretical and
Computational Physics, The University of Hong Kong, Pokfulam Road,
Hong Kong, China}

\author{Zheng-Yuan Xue}
\affiliation{Laboratory of Quantum Information Technology and
SPTE, South China Normal University, Guangzhou, China}

\author{Hui Yan}
\affiliation{Laboratory of Quantum Information Technology and
SPTE, South China Normal University, Guangzhou, China}

\author{Z. D. Wang}
\email{zwang@hku.hk}\affiliation{Department of Physics and Center of Theoretical and
Computational Physics, The University of Hong Kong, Pokfulam Road,
Hong Kong, China}

\author{Shi-Liang Zhu}
\email{slzhu@scnu.edu.cn} \affiliation{Laboratory of Quantum
Information Technology and SPTE, South China Normal University,
Guangzhou, China}

\begin{abstract}
We propose an experimental scheme to detect macroscopic Klein
tunneling with spin-orbit coupled Bose-Einstein condensates
(BECs). We show that a nonlinear Dirac equation with tunable
parameters
can be realized with such BECs. 
Through numerical calculations, we demonstrate that macroscopic
Klein tunneling can be clearly detected under realistic
conditions. Macroscopic quantum coherence in such relativistic
tunneling is clarified and a BEC with a negative energy is shown
to be able to transmit transparently through a wide Gaussian
potential barrier.

\end{abstract}
 \maketitle

\section{introduction}
Shortly after the relativistic equation of electron was established 
by Dirac, Klein used it to study electron scattering by a
potential step and found that there exists a nonzero transmission
probability even though the potential height tends to
infinity\cite{Klein1929}, in contrast to the scattering of a
non-relativistic particle. This phenomenon has been referred to as
Klein tunneling (KT). KT is an intrinsic relativistic effect and
is interpreted as a fundamental property of Dirac equation that
particle and
antiparticle states are inherently linked together as two 
components of the same spinor wavefunction \cite{N.
Dombey}.

This unique scattering process has attracted lots of interest over
the past eighty years but failed to be directly tested by elementary
particles due to the requirements of currently unavailable electric
field gradients \cite{M. I. Katsnelson}.
Interestingly, the dynamics of particles in some systems, such as
electrons in graphene \cite{M. I. Katsnelson} and trap ions
\cite{Lamata,Gerritsma} etc., may be described by effective
relativistic wave equations and have been proposed to observe such
relativistic tunneling. Ultra-cold atoms in optical lattices
\cite{Zhu2007} and light-induced gauge fields \cite{Ruseckas} are
also able to behave as relativistic particles
\cite{Vaishnav,zhu2009}. Recent experiments in graphene
heterojunctions \cite{Stander,Young} have provided some indications 
for KT. However, the existence of disorders and interactions in
these solid-state systems makes it hard to realize full ballistic
scatterings. In addition, it seems hard to unambiguously observe
KT in graphene since it is a typical two-dimensional (2D) system,
while scattering in a 2D system is a combination of perfect
transmission for normally incident particles (a relativistic
effect) and exponentially decay tunneling for obliquely incident
particles (a non-relativistic effect). Moreover, KT as well as
Zitterbewegung effect have been experimentally simulated with the
trapped ions\cite{Gerritsma}.

In this paper we propose a feasible experimental scheme to observe
macroscopic KT with spin-orbit coupled BECs \cite{zhaihui,NIST}.
We demonstrate that a one-dimensional nonlinear Dirac equation
(NLDE) with tunable parameters can be realized with a spinor BEC
in the presence of a light-induced gauge field. Through numerical
simulations, we demonstrate that a macroscopic KT can be observed
under realistic conditions. The simple configuration of gauge
field, in combination with controllable dimensions, interactions
and potential barriers may provide us with a clean and tunable
platform for investigation of interesting relativistic tunneling
effects.

 We investigate the relativistic tunneling of a macroscopic
quantum object by comparing the transmission coefficients between
a BEC in the absence of interactions and an incoherent ensemble
average of non-condensed atoms. In addition, we find that a
realistically weak interaction between atoms  slightly affects the
transmission coefficients. The main feature of a BEC is that all
atoms in the BEC are in the same state and in the same phase and
thus the BEC can be considered  a macroscopic object. So the
tunneling of a BEC we study is the coherent scattering of a
macroscopic object. Tunneling in the former shows a distinct
difference in relativistic effects between macroscopic objects and
the ensemble average of some microscopic particles, while KT has
been studied previously only within a single-particle scenario. We
also present another unexpected result: that a BEC with a negative
energy can almost completely transmit through a Gaussian barrier.
Since KT is a relativistic phenomenon associated with an
anti-particle in the potential, our proposed spin-orbit coupled
BEC can mimic a macroscopic 'anti-BEC' (a super-atom made from
'anti-atoms'), at least in a scattering problem. Therefore, the
mimicked 'anti-BEC' may open the possibility of exploring exotic
relativistic effects of a macroscopic body (even for very large
antimatter), in contrast to the conventional wisdom that
relativistic effects are only clearer for a microscopic particle.

The paper is organized as follows. In Sec. II we propose an
approach to realize a spin-orbit coupled atomic gas through a
$\Lambda$-level configuration, and then demonstrate that the
dynamics of the atoms should be described by the NLDE when the
atoms are condensed into a BEC. In Sec. III we show that the
region for KT of a single atom can be reached in experiments. Then
we demonstrate in Sec. IV that KT of BECs can be clearly observed.
We also clarify the macroscopic quantum coherence in such
relativistic tunneling and show that a wide Gaussian potential
barrier is transparent for a BEC with a negative energy. In Sec.
V, we present our discussion and conclusion. In the Appendix, we
briefly  review the numerical method to calculate the transmission
coefficient of a single atom scattered by a Gaussian potential.

\begin{figure}[tbhp]
\includegraphics[width=8cm,height=2.5cm]{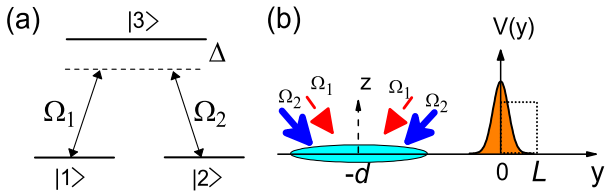}
\caption{(Color online).  Schematic illustration of the system.
(a) Atom with a $\Lambda$-level configuration interacting with
laser beams characterized by Rabi frequencies $\Omega_1$,
$\Omega_2$ and a large detunning $\Delta$. (b) Configuration of
laser beams to realize a Dirac-like equation by the lasers
$\Omega_1$,$\Omega_2$ and an effective Gaussian (square)-shaped
potential induced by another laser beam. Atoms are confined in a
1D waveguide along the $y$ axis and scattered by the potential.}
\end{figure}

\section{realization of a nonlinear Dirac equation with cold atoms}

The Dirac equation with tunable parameters can be realized with
ultracold atoms through two approaches
\cite{Zhu2007,Vaishnav,zhu2009}. Similarly to graphene, it was
proposed that low-energy quasiparticles in a honeycomb optical
lattice should also be described by the relativistic Dirac
equation \cite{Zhu2007}. On the other hand, the Hamiltonian of
cold atoms (without optical lattices) with a certain spin-orbit
coupling, which can be achieved with synthetic gauge fields, is a
Dirac Hamiltonian when the wave number of the atoms is much
smaller than the wave number of the laser beams. It is
demonstrated that the required spin-orbit coupling can be realized
though a tripod level configuration \cite{Vaishnav,zhu2009}. In
this paper we proposed that a $\Lambda$-level configuration is
also feasible for use in the realization of the Dirac equation.

Let us consider the motion of bosonic atoms with mass $m$ in the
y-z plane, with each having a $\Lambda$-level structure
interacting with laser beams as shown in Fig. 1. The ground states
$|1\rangle$ and $|2\rangle$ are coupled to an excited state
$|3\rangle$ through laser beams characterized respectively with
the Rabi frequencies $\Omega_1=\Omega\cos(\kappa_y y)e^{-i\kappa_z
z}$ and $\Omega_2=\Omega\sin(\kappa_y y)e^{i(\pi-\kappa_zz)}$,
where $\Omega=\sqrt{|\Omega_1|^2+|\Omega_2|^2}$. As shown in Fig.1
(b), the Rabi frequencies $\Omega_1$ and $\Omega_2$ can be
realized, respectively, with a pair of lasers
$\Omega_{1\pm}=\frac{1}{2}\Omega \exp[i(-\kappa_z z \pm \kappa_y y
)]$ and $\Omega_{2\pm}=\frac{1}{2}\Omega \exp\{i[-\kappa_z
z\pm(\kappa_y y+ \pi/2)]\}$, where $\kappa_y=\kappa\cos\varphi$
and $\kappa_z=\kappa\sin\varphi$ with $\kappa$ being the wave
number of the lasers and $\varphi$ being the angle between the
laser and the $y$ axis. The Hamiltonian of a single atom reads
 $H=\frac{\mathbf{P}^2}{2m}+V(\mathbf{r})+H_I$,
 where $V(\mathbf{r})=\sum_{j=1}^3(V_T(\mathbf{r})+V_b(\mathbf{r}))|j\rangle\langle
j|$ denotes the full external potentials (including the trapping
potentials $V_T$ and the scattering potential $V_b$) and the
interaction Hamiltonian 
$H_I=\hbar\Delta|3\rangle\langle3| %
-(\sum_{j=1}^2\hbar\Omega_j|3\rangle\langle j|+h.c.)$, with
$\Delta$ as the detuning. Diagonalizing $H_I$ yields the
eigenvalues $\hbar \{[\Delta
-\sqrt{\Delta^2+4\Omega^2}]/2,0,[\Delta+\sqrt{\Delta^2+4\Omega^2}]/2
\}.$  In the large detuning case, the two eigenstates
corresponding to the first two eigenvalues span a near-degenerate
subspace, and can be considered a pseudo-spin with spin-orbit
coupling induced by a gauge potential \cite{Ruseckas,Xiongjun}.
Under this condition we obtain the effective Hamiltonian
\begin{equation}
\label{HS} H=
\frac{p_y^2+p_z^2}{2m}+v_y\sigma_yp_y+v_z\sigma_zp_z+\gamma_z\sigma_z+V_T+V_b,
\end{equation}
where  $v_y=\frac{\hbar \kappa_y}{m}$, $v_z=\frac{\hbar
\kappa_z\Omega^2}{2m\Delta^2}$, and
$\gamma_z=\frac{\hbar^2\Omega^2}{4m\Delta^2}[\kappa_y^2-(1+\Omega^2/\Delta^2)\kappa_z^2]+\frac{\hbar\Omega^2}{2\Delta}$.
In the derivation, we have dropped an irrelevant constant and
assumed that the potentials $V(\mathbf{r})$ are spin-independent.
Furthermore, the atomic gas can well be confined by a 1D optical
waveguide along the $y$ axis \cite{zhu2009}, so we may further
restrict our study in the 1D system. Therefore, both tripod- and
$\Lambda$-level configurations can be used, in principle, in the
realization of the Dirac equation. Compared with the tripod
configuration \cite{Vaishnav,zhu2009}, a large detuning is
necessary in the $\Lambda-$ configuration. However, the laser
beams are simpler in the $\Lambda$-level configuration.
Furthermore, the pseudospins in the $\Lambda-$ configuration would
be more robust against the collision of atoms since they are
constructed by the lowest two dressed states, while the two dark
states in the tripod configuration are not the ground states.

We assume that the interaction can be described by an effective 1D
interacting strength $g=2\hbar^2a_sN/(ml_\bot^2)$, where $a_s$ is
the scattering length, $N$ is the particle number, and $l_\bot$ is
the oscillator length associated with a harmonic vertical
confinement. The interaction between the atoms (per particle)
should be much smaller than the confinement frequency (about
kHz)\cite{Petrov}, and thus is also much smaller than $\Omega$ (in
the MHz range), therefore the interaction can not pump the atoms
outside of the near-degenerate subspace. Under the condition
$p_y\ll \hbar\kappa_y$,
we can safely neglect the $p_y^2$ term. In addition, we assume
that the bosonic atoms are condensed into a BEC state. Within the
Gross-Pitaevskii formalism, the interacting bosons in the
near-degenerate subspace are then effectively described by a 1D
NLDE as  $ i\hbar\partial_t\Psi =H_{ND} \Psi $ \cite{Merkl}, where
\begin{equation}
\label{NLDE} H_{ND}=-i\hbar
v_y\sigma_y\partial_y+\gamma_z\sigma_z+g\Psi^\dag\cdot\Psi+V_T+V_b
\end{equation}
with $v_y$ being the effective speed of light and $\gamma_z$ as
the effective rest energy of the cold atoms. It is a remarkable
feature that all parameters, $v_y$, $\gamma_z$ and $g$, can be
controlled experimentally, providing us with a tunable platform
for exploration of the relativistic quantum effects.

\section{Klein tunneling of a single atom}
We now address the relativistic quantum tunneling that can be
observed with cold atoms. To get an intuitive physics picture, we
first consider a single atom with energy $E$ scattered by a square
potential with width $L$ and potential height $V_s$. Such a
potential can be experimentally formed by a laser beam with a
flat-top profile \cite{Tarallo}. The transmission coefficient
$T_D$ for the so-called KT regime
 $V_{\rm{s}}>E+\gamma_z$ \cite{N. Dombey}, can be
obtained explicitly as
\begin{equation}
\label{T_D} T_D=\left[1+(\eta -\eta^{-1})^2\sin^2(\beta
L)/4\right]^{-1},
\end{equation}
where
$\eta=\sqrt{\frac{(V_{\rm{s}}-E+\gamma_z)(E+\gamma_z)}{(E-V_{\rm{s}}+\gamma_z)(\gamma_z-E)}}$
and
$\beta=\sqrt{\frac{(V_{\rm{s}}-E-\gamma_z)}{(V_{\rm{s}}-E+\gamma_z)}}/\hbar$.
Compared with the well-known property in nonrelativistic quantum
mechanics that the transmission coefficient decreases
mono-exponentially with the height $V_s$ or width $L$, a
distinctly different feature within this KT region is that the
tunneling amplitude is an oscillation function of $V_s$ or $L$
even when the kinetic energy of the incident particle is less than
the height of the barrier. This relativistic effect can be
attributed to the fact that the incident particle in a positive
energy state can propagate inside the barrier by occupying a
negative energy state, which is also a plane wave aligned in
energy with that of the particle continuum outside. Matching
between positive and negative energy states across the barrier
leads to high-probability tunneling. We take the atoms of $^{7}$Li
as an example. If we choose the practical parameters
$\kappa_y=10^7$
$\rm{m^{-1}}$, $\kappa_z=0.8\times10^7$ $\rm{m^{-1}}$, 
$\Omega=10^7$ Hz and 
$\Delta=10^9$ Hz, it is found that the Klein regime corresponds to
the Rabi frequency $\Omega_b^{\rm{s}}>0.162$ MHz, which can be
easily achieved in experiments. So we have demonstrated from a
simple example that it is feasible to observe KT with cold atoms.

\section{Klein tunneling of atomic condensates}
As for a practical experiment it is required to release two
conditions: the trajectory of a single atom is hard to detect, and
it is much easier  to measure the density evolution of an ensemble
of atoms in experiments. Compared with the square potential,
 a Gaussian potential $V^{\rm{G}}_b(y,\nu)=\nu
V_{\rm{G}}e^{-y^2/\sigma^2}$, where $V_{\rm{G}}$ is the height and
$\sigma$ characterizes the spatial variance, is much easier to be
generated. Here $\nu$ donates a barrier $(\nu=+)$ or a potential
well $(\nu=-)$, and the potential barrier (well) can be realized
by focusing a blue- (red-) detuned far-off-resonant
Gaussian-shaped laser beam. However, the conditions of resonant
transmission vary with the velocity and the width of the
potential, and thus both the ensemble of atoms and the Gaussian
potential may smooth the oscillations in the transmission
coefficient. So it is natural to ask whether KT can still be
observed in an ensemble of atoms. Surprisingly we illustrate below
that KT of a BEC may be observed very clearly.

\begin{figure}[tbhp]
\label{BECscatter}
\includegraphics[width=8cm,height=3cm]{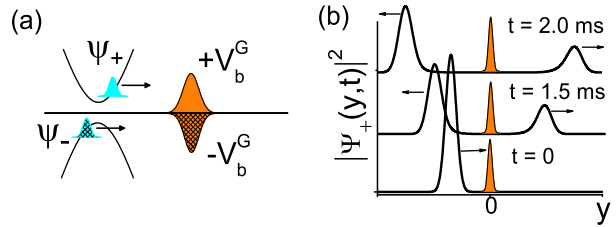}
\caption{ (Color online). (a) A schematic diagram showing four
kinds of scattering events. (b) Normalized density distribution in
a scattering process at time $t=0,\ 1.5 \ \text{and} \ 2.0\
\text{ms}$. The peaks at $y=0$ are the Gaussian barriers.}
\end{figure}

We assume that a BEC consisting of $^7$Li is initially trapped in
a harmonic trap which moves along the $y$ axis. 
At the initial time $t=0$, the center of the trap is located at
$y=-d$, and the center of the Gaussian potential is at $y=0$. The
trap is turned off at $t=0$ and then we calculate the evolution of
the density profile of the atomic gas after a long enough time for
scattering. The single-atom dispersion described in Eq.
(\ref{NLDE}) is characterized by two branches $E_\pm(k_y)= \pm
(\gamma_z^2+\hbar^2v_y^2k_y^2)^{1/2}$, where the lower (upper)
branch represents the negative (positive) energy state. One can
prepare an initial BEC with a designated mode $k_0$ at the
positive or negative energy branch.
The two branches allow us to study a more fruitful tunneling
problem: there are four classes of scattering which describe the
wave function  $\Psi_\mu$ [$\mu (= \pm)$] scattered by the
potential $V_b^G(y,\nu)$, as shown in Fig. 2(a).

The BEC in a harmonic trap can be well described by a Gaussian wave
packet, so we may choose the initial wave function as
\begin{equation}
\label{Wavefunction}
\Psi_\mu(y,0)=\frac{1}{\sqrt{
l_0\sqrt{\pi}}}e^{i\mu k_0y}e^{-(y+d)^2/2 l_0^2}\phi_\mu,
\end{equation}
where $l_0$ is the width, $k_0$ is the central wave number of the
wave-packet and the spinors $\phi_\mu$ are defined as $ \phi_+ = (
i\cos\xi, -\sin\xi)^{T},$ $\phi_- = (-i\sin\xi, \cos\xi )^{Tr}$
with $\xi=\frac{1}{2}\arctan(\hbar v_yk_0/\gamma_z)$ and $Tr$ the
transposition of matrix. This wave function describes a Gaussian
wave packet with the central velocity $\hbar (\kappa_y+\mu k_0
)/m$ moving along the $y$-axis. After evolution governed by the
Dirac-type Eq. (\ref{NLDE}) with time $t$, the finial wave
function becomes
\begin{equation}
\label{PsiEvolution} \Psi_\mu(y,t)=\hat{\mathcal{T}} \exp\left(
-\frac{i}{\hbar}\int_0^t H_{ ND} dt\right) \Psi_\mu(y,0),
\end{equation}
where $\hat{\mathcal{T}}$ denotes the time ordering operator. We
numerically calculate $\Psi_\mu(y,t)$ in Eq.(\ref{PsiEvolution})
by using the standard split-operator method.
  According to the method of \cite{Larson}, Eq.
(\ref{PsiEvolution}) can be rewritten as
\begin{equation}
\begin{array}{ll}
\Psi_\mu(y,t+\delta t)=\left\{e^{-\frac{i}{2\hbar} v_y\sigma_yp_y
\delta
t}e^{-\frac{i}{\hbar}\gamma_z\sigma_z\delta t}\right.\\
~~~~~~~~~~~~~~~~~~~\times
e^{-\frac{i}{\hbar}\left[V^{\rm{G}}_b(y,\nu)+g|\Psi_\mu(y,t)|^2\right]\delta
t}\\
~~~~~~~~~~~~~~~~~~~\left.\times e^{-\frac{i}{2\hbar} v_y\sigma_yp_y
\delta t}+\mathcal {O}({\delta t}^3)\right\}\Psi_\mu(y,t).
\end{array}
\end{equation}
In the sufficiently short time step $\delta t$, the high-order
term $\mathcal {O}({\delta t}^3)$ (due to the non-commuting) can
be safely neglected. Combining with the Fourier transform between
the position and momentum spaces, we can finally get the numerical
solution of $\Psi_{\mu}(y,t)$ following the computation procedure
step by step with time step $\delta t$.

We have numerically calculated $\Psi_\mu(y,t)$, and found the
existence of stationary solution for the scattering process, with
an example being shown in Fig. 2(b). After tunneling, the incident
wave packet divides into left- and right-traveling wave packets,
and only the latter one is on the transmission side of the
barrier. Thus we can define the transmission coefficient of the
incident wave packet $\Psi_\mu(y,0)$ scattering by a potential
$V_b^G(y,\nu)$ as
\begin{equation}
\label{Tmn} T_{\mu\nu} =
 \int_{\sigma}^\infty
\Psi_\mu^\dag(y,\tau)\Psi_\mu(y,\tau)dy.
\end{equation}
Here $\tau$ (being slightly larger than $d/v_0$) represents a
typical time that the reflected and transmitted wave packets are
sufficiently away from the Gaussian potential.  One can directly
measure the transmission coefficient in Eq. (\ref{Tmn}) since the
spatial density distribution $\rho_\mu
(y,\tau)=|\Psi_\mu(y,\tau)|^2$ can be detected using absorption
imaging \cite{Khaykovich}.

We first look into the tunneling phenomena for a BEC in the
absence of interactions ($g=0$).
We note that there are
two identities $T_{++}=T_{--}$ and $T_{-+}=T_{+-}$ 
since  Eq.(\ref{NLDE}) with $g=0$ is invariant under the charge
conjugation \cite{Norman Dombey}. We plot the transmission
coefficient  $T_{++}$ as a function of the height $V_{\rm{G}}$ and
width $\sigma$ in Fig. 3(a) with the practical parameters. It is
interesting to note that the transmission coefficient decreases
exponentially to zero with $V_{\rm{G}}$ when
$V_{\rm{G}}<V_{\rm{G}}^K$, while it increases and then is an
oscillating function in the Klein region
$V_{\rm{G}}>V_{\rm{G}}^K$, these results are similar to the
results of Eq. (\ref{T_D}) for the square barrier. Here the
critical value of the potential height may  be estimated
approximately using the square barrier with $V_{\rm{G}}^K=
E(k_0)+\gamma_z \approx 0.09$ MHz. Moreover, the feature
$T_{++}=T_{--}$ is also confirmed in the inset in Fig. 3(a). As
for the transmission coefficient $T_{++}(\sigma)$, we may obtain
several tunneling oscillations with the potential width, but it
decreases to 0 when the width is further increased. Although the
amplitude of tunneling oscillation is less than the unit compared
with the tunneling of a single atom, the amplitude of tunneling
oscillation can be more than $0.5$ and meanwhile the period can be
a few micrometers, which is experimentally detectable.

Another interesting feature induced by relativistic effects is
that, a BEC with negative energy can almost completely transmit a
wide Gaussian potential barrier, as shown in Fig. 3(b). The
transmission coefficient $T_{-+}$ is an oscillating function of
the potential width $\sigma$ when $\sigma$ is smaller than $3\
\mu$m, while it saturates quickly to  the unit when the potential
width is larger than $3\ \mu$m, leading to the unexpected result
that a wide Gaussian potential barrier is actually totally
transparent for a BEC. This phenomenon can be understood through
the fact that this scattering feature is actually equivalent to
that of a BEC of positive energy scattered by a Gaussian potential
well because of $T_{-+}=T_{+-}$. We also calculate the
transmission coefficient for the central mode of the wave packet, 
as shown in the inset in Fig. 3(b), which further confirms that a
wide enough Gaussian potential well is transparent. The reason
lies in the fact that, in contrast to the periodic function
(without a saturation value) in a square potential well, the
Gaussian potential well is smooth in the whole space, and thus can
even support adiabatic motions of wave packets in the large width
limit.

\begin{figure}[tbhp]
\label{BECFig}
\includegraphics[width=8.8cm,height=4cm]{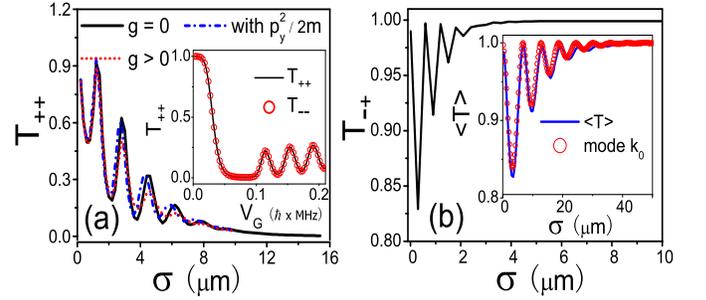}
\caption{ (Color online). KT of  BECs. (a) $T_{++}(\sigma)$ for
$V_{\rm{G}}/\hbar=0.2$ MHz and $T_{++}(V_G)$ (inset) for
$\sigma=5$ $\mu$m. The tunnelings of a BEC with the classic
kinetic energy term and the conventional atomic interaction ($N =
2 \times 10^4$, $l_\perp = 1.4$ $\mu$m and $a_s = 5a_0$ with $a_0$
being the Bohr radius) are also depicted. (b) Coefficients
$T_{-+}(\sigma)$, $\langle T(\sigma)\rangle$ (insert) of one atom
with central mode $k_0$, and of $10^4$ atoms for
$V_{\rm{G}}/\hbar=0.2$ MHz. The other parameters in (a) and (b)
are $l_0=10$ $\mu$m, $k_0=5.5\times10^5$ $\rm{m^{-1}}$,
$\gamma_z/\hbar = 30$ kHz, and $d=4(l_0+\sigma)$. }
\end{figure}

\begin{figure}[tbhp]
\label{BECFig}
\includegraphics[width=8.8cm,height=4cm]{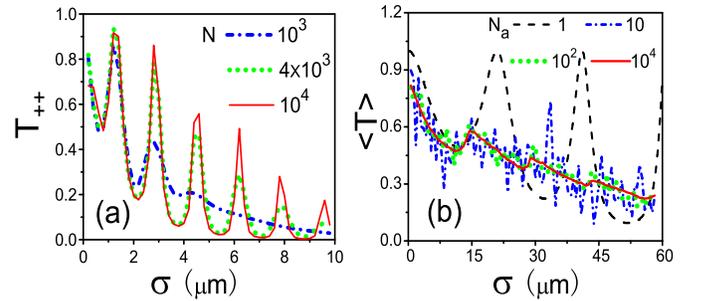}
\caption{ (Color online). Comparison of KT of BECs with that of an
ensemble of non-condensed atoms. (a) $T_{++}(\sigma)$ with
$N=10^3$, $4\times10^3$, $10^4$ are shown by fixing the energy
$E_{int}$ for $l_0=10$ $\mu$m and $d=4(l_0+\sigma)$. (b) $\langle
T(\sigma)\rangle$ with $N_a=1$, $10$, $10^2$, and $10^4$ atoms for
$\sigma_k=5\times10^5$ $\rm{m^{-1}}$. The other parameters in (a)
and (b) are $V_{\rm{G}}/\hbar=0.2$ MHz, $k_0=5.5\times10^5$
$\rm{m^{-1}}$, and $\gamma_z/\hbar = 30$ kHz.}
\end{figure}

The tunneling properties exhibited in Fig. 3(a) and 3(b) are
intrinsic relativistic and macroscopic quantum phenomena that can
not be explained with an incoherent ensemble average of many
atoms. To clarify this point, we calculate the average
transmission coefficients for an ensemble of $N_a$ noninteracting
atoms defined as
\begin{equation}\langle
\label{T_k} T\rangle=\frac{1}{N_a}\sum_{i=1}^{N_a}T(k_i),
\end{equation}
where $T(k_i)$ donates the transmission coefficient for atom $i$
with the wave number $k_i$ scattered by the potential. The
numerical
calculation method for $T(k_i)$ is given in the Appendix. Here we choose 
$k_i$ to be the same Gaussian distribution as that of the initial
BEC wave function $\Psi_\mu (y,0)$, i.e., $k_i \sim
N(k_0,\sigma_k^2)$ with the variance $\sigma_k=1/l_0$. The
$\langle T \rangle$ of $10^4$ atoms is shown in the inset in Fig.
3(b), which is almost the same as that of a single atom since
$\sigma_k$ is small. The differences between $\langle T \rangle $
and $T_{-+}$ in Fig. 3(b)  demonstrate that the tunneling of BEC
is not equivalent to an ensemble average of the individual atoms
even with the same distribution of wave number. The coefficient
$\langle T \rangle$ represents an incoherent transmission of the
individual particles since it is a sum of the transmission
coefficients of all particles. In contrast, the phases of all
atoms in the BEC are the same and then the transmission of a BEC
is coherent. The coherent transmission in a BEC and incoherence in
$\langle T \rangle$ cause the difference in Fig. 3. All particles
in the BEC are in the same phase because the macroscopic number of
particles are condensed in the same state, so the coherent
transmission of the BEC may be called macroscopic quantum
tunneling.

To clarify further the macroscopic quantum phenomena in the
relativistic tunneling of a BEC, we compare the scaling properties
of transmission coefficients for a weakly interacting BEC and an
incoherent ensemble average of atoms. An example of scaling of
$T_{++}$ is plotted in Fig. 4(a).
In the calculations, we have fixed the weak interatomic
interaction energy $E_{int}\approx g/l_0$  and kept the parameter
$\gamma=mgl_0/N\hbar^2\ll1$ \cite{Petrov} for $l_0=5$ $\mu$m when
$N=10^3$, $l_\perp = 1.4$ $\mu$m and $a_s = 5a_0$ ($\gamma\sim
10^{-3}$), both of which restrict our discussions in the regime
for 1D BECs, where Dirac dynamics instead of nonlinear dynamics
dominates. In this case, the increase in particle number is
achieved by proportionally increasing the length $l_0$ of the BEC
with small $\gamma$. For comparison, we also calculate the scaling
of $\langle T\rangle$ for the atom numbers of 1 (with $k_i=k_0$),
$10$, $10^2$, and $10^4$ in Fig. 4(b). Comparing Fig. 4(a) with
Fig. 4(b), a distinct difference between the BEC and  the ensemble
average of the individual atoms is that, the coefficient $T_{++}$
increases  with increasing atomic number of the BEC, while the
coefficient $\langle T\rangle$ decreases with the increasing of
the atomic number. However, in order to keep the same interaction
parameter in the above calculation, we have increased
simultaneously the particle number and the width of the Gaussian
wave packet. In this way the momentum distribution of the wave
functions is shrunk,  which is a dominant reason for the above
scaling feature. That many atoms may condense into the same
momentum state is essential for the observation of KT in a BEC.

\section{discussion and conclusion}
Before concluding, we wish to make two additional comments. (i) 
To judge the feasibility of the Dirac approximation in
Eq.(\ref{NLDE}), the coefficients $T_{++}$ with or without the
quadratic term are compared in Fig. 3(a). It is shown that the
quadratic term leads to merely a slight left-shift of the
tunneling peaks. This phenomenon can be interpreted by the fact
that the wavelength of the BEC inside the barrier decreases
slightly in the presence of the additional low kinetic energy.
This result verifies that the approximation leading to the Dirac
equation is well satisfied.  (ii) In Fig. 3(a), we have also
calculated the transmission coefficient for BECs with conventional
atomic interactions without Feshbach resonance, in which case the
experimental setup can be simplified. The result shows that the
effect of the realistically weak interaction is small; it merely
smooths the tunneling oscillation slightly. Therefore the exotic
tunneling phenomena addressed here survive in the case of weak
interaction between atoms.

In summary, we have proposed an experimental scheme to detect
macroscopic KT using a spin-orbit-coupled BEC. Through numerical
simulations, we have elaborated that such macroscopic KT can be
observed under realistic conditions. In view of  the fact that a
spin-orbit-coupled BEC was realized in a very recent experiment
\cite{NIST}, it is anticipated that the present proposal will be
tested in an experiment in the near-future.

\section{acknowledgments}
This work was supported by the NSFC (Nos. 11125417 and 10974059),
the SKPBR (No.2011CB922104), the GRF and CRF of the RGC of Hong
Kong.

\vspace{0.5cm} \textbf{APPENDIX: The derivation of $T(k_i)$ in
Eq.(\ref{T_k})}

\vspace{0.3cm}

The Dirac equation for particle scattering by a Gaussian potential
can not be solved analytically for an incoming atom with energy
$E_i=\sqrt{(\hbar v_y k_i)^2+\gamma_z^2}$ and momentum $p_i=\hbar
k_i$. However, here we adopt an efficient method to solve it
numerically based on transfer matrix methods \cite{zhu2009}. The
numerical procedures are outlined as follows. First, one cuts the
Gaussian potential into a spatially finite range $y\in[-y_c,y_c]$,
where the cutoff position $y_c$ should be chosen to guarantee that
the potential height outside the range is low enough to be
transparent for the atoms, i.e., $V_b^{\rm{G}}(y_c)\ll E_i, V_{\rm
G}$. Second, one  divides this range equally into $n$ spindly
segments, and each segment may be considered a square potential if
$n$ is large enough. The potential height of the $j$-th ($j=1, 2,
\cdots, n$) square potential is given by
$V_j=V^{\rm{G}}_b(y_j+f/2)$ with $y_j=-y_c+(j-1)f$ and the width
of each potential $f=2y_c/n$. In this case, the Gaussian potential
can be viewed approximately  as a sequence of connective small
square potential barriers, and thus the transmission coefficient
$T(k_i)\approx 1/|m_{11}|^2$, where $m_{11}$ is the first element
in the whole transfer matrix $M=M_nM_{n-1}\cdots M_j\cdots
M_2M_1$. Here $M_j$ denotes the transfer matrix of the $j$-th
square potential barrier, whose explicit elements are given by
\cite{zhu2009}
\begin{equation}
\begin{array}{ll}
\label{Element_D}
(M_j)_{11}=(\cos\frac{p_j f}{\hbar}+i\frac{\kappa^2+\kappa_j^2}{2\kappa\kappa_j}\sin\frac{p_j f}{\hbar})e^{-\frac{i}{\hbar}p_i f},\\
(M_j)_{12}=(i\frac{\kappa_j^2-\kappa^2}{2\kappa\kappa_j}\sin\frac{p_j f}{\hbar})e^{-\frac{i}{\hbar}p_i(y_j+y_{j+1})}, \\
(M_j)_{21}=(M_j)_{12}^*, \\    (M_j)_{22}=(M_j)_{11}^*,
\end{array}
\end{equation}
where $\kappa=(E_i-\gamma_z)/(v_y p_i)$ and
$\kappa_j=(E_i-\gamma_z-V_j)/(v_y p_j)$ with
$(E_i-V_j)^2=v_y^2p_j^2+\gamma_z^2$. Note that this numerical
calculation scheme recovers the non-relativistic scattering
governed by the Schr\"{o}dinger equation.

\end{document}